\documentclass[prd,superscriptaddress,twocolumn,amsfonts,amssymb,amsmath,showpacs]{revtex4-2}
\usepackage{bm}
\usepackage{amsfonts}
\usepackage{latexsym}
\usepackage[latin1]{inputenc}
\usepackage{graphicx}
\usepackage{amsmath}
\usepackage{palatino}
\usepackage{caption}
\usepackage{subcaption}
\usepackage{mathpazo}
\usepackage{textcomp}
\usepackage{epsfig}
\usepackage{float}
\usepackage{booktabs}
\usepackage{dcolumn}
\usepackage{booktabs}
\usepackage{multirow}
\usepackage{hyperref}
\hypersetup{colorlinks,citecolor=blue}
\usepackage{amsmath}
\usepackage{xcolor}
\usepackage{orcidlink}
\usepackage{commath}

\def\jnl@style{\it}
\def\aaref@jnl#1{{\jnl@style#1}}

\def\aaref@jnl#1{{\jnl@style#1}}

\def\aj{\aaref@jnl{AJ}}                   
\def\apj{\aaref@jnl{ApJ}}                 
\def\apjl{\aaref@jnl{ApJ}}                
\def\apjs{\aaref@jnl{ApJS}}               
\def\apss{\aaref@jnl{Ap\&SS}}             
\def\aap{\aaref@jnl{A\&A}}                
\def\aapr{\aaref@jnl{A\&A~Rev.}}          
\def\aaps{\aaref@jnl{A\&AS}}              
\def\mnras{\aaref@jnl{Mon.~Not.~Roy.~Astron.~Soc.}}             
\def\prd{\aaref@jnl{Phys.~Rev.~D}}        
\def\prc{\aaref@jnl{Phys.~Rev.~C}}  
\def\prl{\aaref@jnl{Phys.~Rev.~Lett.}}    
\def\qjras{\aaref@jnl{QJRAS}}             
\def\skytel{\aaref@jnl{S\&T}}             
\def\ssr{\aaref@jnl{Space~Sci.~Rev.}}     
\def\zap{\aaref@jnl{ZAp}}                 
\def\nat{\aaref@jnl{Nature}}              
\def\aplett{\aaref@jnl{Astrophys.~Lett.}} 
\def\apspr{\aaref@jnl{Astrophys.~Space~Phys.~Res.}} 
\def\physrep{\aaref@jnl{Phys.~Rep.}}      
\def\physscr{\aaref@jnl{Phys.~Scr}}       
\def\commat{\aaref@jnl{Comm.~Math.~Phys.}}              
\def\science{\aaref@jnl{Science}}               
\def\cqg{\aaref@jnl{Classical Quant.~Grav.}}            
\def\jpcs{\aaref@jnl{JPCS}}                                     
\def\ijmpd{\aaref@jnl{Int.~J.~Mod.~Phys.~D}}                    
\def\grg{\aaref@jnl{Gen.~Relat.~Gravit.}}               
\def\rpp{\aaref@jnl{Rep.~Prog.~Phys.}}          
\def\npa{\aaref@jnl{Nucl.~Phys.~A}}        
\def\lrr{\aaref@jnl{Living Rev.~Rel.}}                   
\def\jcap{\aaref@jnl{J.~Cosmology Astropart.~Phys.}}    
\def\rmp{\aaref@jnl{Rev.~Mod.~Phys.}}   
\def\epjc{\aaref@jnl{Eur.~Phys.~J.~C}}

\allowdisplaybreaks[1]

\begin{document}

\color{black}       
\title{\bf  The scalar-torsion gravity corrections in the first-order inflationary models}

\author{I. V. Fomin\orcidlink{0000-0003-1527-914X}}
\email{ingvor@inbox.ru}
\affiliation{Bauman Moscow State Technical University, Russia}

\author{S. V. Chervon\orcidlink{0000-0001-8898-3694}}
\email{chervon.sergey@gmail.com}
\affiliation{Bauman Moscow State Technical University, Russia}

\author{L.K. Duchaniya \orcidlink{0000-0001-6457-2225}}
\email{duchaniya98@gmail.com}
\affiliation{Department of Mathematics,
Birla Institute of Technology and Science-Pilani, Hyderabad Campus, Hyderabad-500078, India.}

\author{B. Mishra\orcidlink{0000-0001-5527-3565}}
\email{bivu@hyderabad.bits-pilani.ac.in}
\affiliation{Department of Mathematics, Birla Institute of Technology and Science-Pilani, Hyderabad Campus, Hyderabad-500078, India.}

\begin{abstract}
\textbf{Abstract}:
The corrections to the cosmological models induced by non-minimal coupling between scalar field and torsion are considered. To determine these corrections in explicit form, the power-law parametrization of these corrections are proposed. The estimates of possible influence of non-minimal coupling between scalar field and torsion on cosmological parameters for inflationary models implying linear relation between tensor-to-scalar ratio and spectral index of scalar perturbations are obtained. A procedure for verifying these inflationary models due to the observational constraints on the values of cosmological perturbation parameters is also considered.

\end{abstract}

\maketitle

\section{Introduction}\label{SEC1}

In the current cosmological models, a crucial aspect is the characterization of two phases of accelerated expansion: the initial inflationary era in the early Universe~\cite{Baumann:2014nda, Chervon:2019sey} and the accelerated period in the present cosmic era~\cite{Steinhardt:2008nk}.
To describe the dynamics of the expansion of the Universe at different stages of its evolution and observed large-scale structure of the Universe, general relativity \cite{Baumann:2014nda, Chervon:2019sey} and various modified theories of gravity are considered \cite{Clifton:2011jh, Nojiri:2017ncd, Ishak:2018his, Shankaranarayanan:2022wbx}. Teleparallel equivalent of general relativity (TEGR) is another geometrical approach, where torsion scalar $T$ is considered instead of scalar curvature $R$~\cite{deAndrade:2000kr, Arcos:2004tzt,Geng:2011aj, Myrzakulov_2015}. The different modifications of TEGR are considered to construct and analyze the cosmological models both at the level of background dynamics~\cite{Bamba:2012vg,Bamba:2013jqa,
Bamba:2014zra,Skugoreva:2014ena, Jarv:2015odu, Awad:2017ign, Krssak:2018ywd, Hohmann:2018rwf,Gonzalez-Espinoza:2020jss,Jarv:2021ehj,Duchaniya_2023tphi, Duchaniya_2023noet, Kadam:2022lgq,Paliathanasis:2021gfq,Paliathanasis:2022xoq,Leon:2022oyy,Duchaniya:2024vvc,STFI-CHaadaeva} and with taking into account evolution of cosmological perturbations~\cite{Chen:2010va, Golovnev:2018wbh, Raatikainen:2019qey,
Gonzalez-Espinoza:2019ajd,Gonzalez-Espinoza:2020azh,Chervon:2023gio,STFI-Fomin-2024} as well.
We can also note the possibility of constructing the cosmological models based on braneworld teleparallel gravity ~\cite{Nozari:2012qi,Nozari:2012qi,Geng:2014yya,Behboodi:2014tda,Behboodi:2015uqa,Moreira:2021fva}.
It should be noted that cosmological models including scalar field as the source of accelerated expansion of the Universe can be considered with taking into account the Galilean field self-interaction~\cite{Gonzalez-Espinoza:2019ajd} and without this term~\cite{Gonzalez-Espinoza:2020azh} as well.

In $f(T)$ gravity, equations of cosmological dynamics are of second order as in the case of general relativity(GR). However, instead of using the curvature defined through the Levi-Civita connection, in case of $f(T)$ gravity the Weitzenb\"ock connection is used, which implies zero curvature and non-zero torsion~\cite{Cai:2015emx, Duchaniya:2022rqu, Duchaniya_2024ab}. The fact that the dynamic equations for $f(T)$ gravity are of second order~\cite{Cai:2015emx}, it makes these theories simpler than those modified by curvature invariants~(see, for example, in \cite{Clifton:2011jh,Nojiri:2017ncd,Ishak:2018his,Shankaranarayanan:2022wbx}). This type of cosmological model is of considerable interest in theoretical cosmology. It should also be noted that the speed of gravitational waves for the case of $f(T)$ gravity is equal to the speed of light in vacuum, which is similar to both the case of GR and scalar-tensor theories of gravity and satisfies the restrictions on the speed of gravitational waves from astrophysical objects according to the current cosmological observations~\cite{Li:2018ixg}.

It should also be noted that scalar-torsion gravity theories are also used to construct cosmological models of the current stage of the evolution of the Universe~\cite{Kadam:2022lgq,Paliathanasis:2021gfq,Paliathanasis:2022xoq,Leon:2022oyy,Duchaniya:2024vvc,Rodriguez-Benites:2024pce}. Currently, various possibilities for modifying the standard $\Lambda$CDM-model are being considered to alleviate $H_0$ and $\sigma_8$ tension in the $\Lambda$CDM-model (see, Refs.~\cite{DiValentino:2018gcu,Riess:2020fzl,DiValentino:2020zio,DESI:2024aqx,Vagnozzi_2020newph, Vagnozzi_2023, Herold:2023prd, Poulin:2022sgp, Poulin:2023lkg}). The models based on the scalar-torsion gravity including the Galilean field self-interaction~\cite{Duchaniya:2024vvc} and without this term~\cite{Rodriguez-Benites:2024pce} can be considered to modify the $\Lambda$CDM-model.

Thus, the teleparallel equivalent of general relativity (TEGR) and its modifications are the basis for the construction of cosmological models which can be  verifiable by observational data, that corresponds to the growing interest in this type of gravity models and their physical applications in current cosmology~   \cite{Skugoreva:2014ena,Jarv:2015odu,Awad:2017ign,Krssak:2018ywd,Hohmann:2018rwf,Chen:2010va,Golovnev:2018wbh,Raatikainen:2019qey,
Gonzalez-Espinoza:2019ajd,Gonzalez-Espinoza:2020azh,Gonzalez-Espinoza:2020jss,Jarv:2021ehj,  Duchaniya_2023tphi, Duchaniya_2023noet, Kadam:2022lgq}.
When constructing and analyzing cosmological models based on TEGR modifications, the following questions arise: how do these modifications affect cosmological parameters and what possible physical effects do they correspond to? To answer these questions, the approach associated with a model-independent assessment of the influence of modified theories of gravity on cosmological parameters seems promising. In \cite{Fomin:2017sbt,Fomin:2020woj,Fomin:2022ozv,Fomin:2020caa,Fomin:2017vae,Fomin:2017qta,Fomin:2018typ,Fomin:2020hfh} methods were proposed for analyzing cosmological models with modified gravity theories (scalar-tensor gravity and Einstein-Gauss-Bonnet gravity) based on functional and parametric connections to the Einstein gravity in the Friedman-Robertson-Walker space-time. This approach made it possible to directly assess the influence of these modifications of the Einstein gravity on the background dynamics and parameters of cosmological perturbations. In~\cite{Chervon:2023gio} this approach was used for the reconstruction of the scalar-torsion gravity theories from the physical potentials of a scalar field. Also, in~\cite{STFI-CHaadaeva,STFI-Fomin-2024} the cosmological models

In this work, we present a model-independent approach to assess the impact of the non-minimal coupling between the scalar field and torsion on cosmological parameters during the inflationary phase of the Universe evolution. In this approach, the influence can be estimated explicitly for cosmological dynamics, scalar field potential and cosmological perturbations parameters. The proposed power-law relationship between coupling function and Hubble parameter allows us to reduce the assessment of such influence to the estimation of the possible values of the velocity of the scalar perturbation.
This feature of the proposed approach distinguishes it from previously considered methods of analyzing cosmological inflation models based on scalar-torsion gravity theories (see, Refs.~\cite{Gonzalez-Espinoza:2019ajd,Gonzalez-Espinoza:2020azh,Gonzalez-Espinoza:2020jss,Chervon:2023gio,STFI-CHaadaeva,STFI-Fomin-2024}).
The corrections inspired by non-minimal coupling between scalar field and torsion lead to the verification of the inflationary models implying a linear relation between tensor-to-scalar ratio and spectral tilt of scalar perturbations by the observational constraints.

The paper is organized as follows: In Sec-\ref{SEC2}, we have considered the dynamical equations for cosmological models based on the scalar-torsion gravity.  Sec-\ref{SEC3} proposes a power-law relation between the non-minimal coupling function and the Hubble parameter $F=(H/\lambda)^{n}$. The implications of this relation are also discussed. In Sec-\ref{SEC4}, we have obtained the parameter $\lambda$ in explicit form based on the conditions for the correct completion of the inflationary stage. Sec-\ref{SEC5} provides restrictions on the values of parameter $n$ and on the possible types of inflationary models. Also, it was shown that reducing the cosmological models under consideration to the case of $f(T)$-gravity leads to the pure de Sitter stage.
In Sec-\ref{SEC6} we formulate conditions on the slow-roll parameters and condition of the exit from inflation, which restrict the possible types of inflationary models.
In Sec-\ref{SEC7}, it has been shown that the considered inflationary models with non-minimal coupling between scalar field and torsion can be verified from observational data, unlike models with minimal coupling. Estimates of the propagation velocity of scalar perturbations for the verified inflationary models are also given. A summary of our investigations has been presented in Sec. \ref{SEC8}.

\section{Dynamical equations for  cosmological models based on the scalar-torsion gravity}\label{SEC2}

The action of generalized scalar-torsion theory with self-interaction of Galilean-like field in the system of units $M^2_{p}=(8\pi G)^{-1}=c=1$ can be defined as~\cite{Gonzalez-Espinoza:2019ajd},
\begin{eqnarray}
&&S=\int d^4x e \left[\frac{1}{2}F(\phi)T+P(\phi,X)-\square \phi G(\phi,X)\right]\,,
\label{act}
\end{eqnarray}

where $F(\phi)$ is the function of the scalar field $\phi$; $P(\phi,X)$, $G(\phi,X)$ are functions of the scalar field and its kinetic energy $X=-\frac{1}{2}\phi_{,\mu}\phi^{,\mu}$ and $\phi_{,\mu}=\partial_{\mu}\phi=\frac{\partial\phi}{\partial x^{\mu}}$. $T$ be the torsion scalar, $e^{A}_{\mu}$ are the tetrad components in a coordinate basis, $e=det(e^A_{\mu})=\sqrt{-g}$ and $G(\phi,X)$ is the Galileon-type field self-interaction.
Also, the function $P$ can be considered as the function of the potential of a scalar field and its kinetic energy $P=P\left(V(\phi), X\right)$.

The torsion scalar is defined as follows~\cite{Gonzalez-Espinoza:2019ajd,Gonzalez-Espinoza:2020azh,Gonzalez-Espinoza:2020jss}
\begin{eqnarray}
\label{T}
&&T= S_{\gamma}^{~\alpha\beta}\,T^{\gamma}_{~\alpha\beta}\,,\\
\label{TT}
&&T^{\gamma}_{~\alpha\beta}= e_{A}^{~\gamma}\left[\partial_{\alpha}e^{A}_{~\beta}
 -\partial_{~\alpha}e^{A}_{~\beta}+\omega^{A}_{~B\alpha}\,e^{B}_{~\beta}
 -\omega^{A}_{~B\beta}\,e^{B}_{~\alpha}\right]\,,\\
 \label{TS}
&& S_{\gamma}^{~\alpha\beta}=\frac{1}{2}\left(K^{\alpha\beta}_{~~\gamma}+\delta^{\alpha}_{~\gamma} \,T^{\mu\beta}_{~~\mu}-\delta^{\beta}_{~\gamma}\,T^{\mu\alpha}_{~~\mu}\right)\,,\\
\label{TK}
&&  K^{\alpha\beta}_{~~\gamma}= -\frac{1}{2}\left(T^{\alpha\beta}_{~~\gamma}
 -T^{\alpha\beta}_{~~\gamma}-T_{\gamma}^{~\alpha\beta}\right)\,,
\end{eqnarray}

where $T^{\gamma}_{~\alpha\beta}$ is the torsion tensor, $S_{\gamma}^{~\alpha\beta}$ is the super-potential, $ K^{\alpha\beta}_{~~\gamma}$ is the contorsion tensor, and spin connection $\omega^{A}_{~B \alpha}=0$ for a special class of inertial frames.

We consider a special case of negligible self-interaction of a scalar field as,
\begin{eqnarray}
\label{GN}
&&\frac{1}{2}F(\phi)T\gg \square \phi G(\phi,X).
\end{eqnarray}

So, the action  Eq.~(\ref{act}) is reduced to the following form
\begin{eqnarray}
&&S\approx\int  d^4x e \left[\frac{1}{2}F(\phi)T+P(\phi,X) \right].
\label{act1}
\end{eqnarray}

In~\cite{Gonzalez-Espinoza:2019ajd}, the different inflationary models corresponding to assumption (\ref{GN}) and with non-negligible Galileon field self-interaction were considered.
Also, the verifiable models of the current stage of the Universe's evolution can be constructed with non-negligible Galileon self-interaction~\cite{Duchaniya:2024vvc} and without this term~\cite{Rodriguez-Benites:2024pce} as well.
Thus, we will use assumption (\ref{GN}) to consider the need to include the Galilean field self-interaction in actual cosmological models.

Thus, we shall consider the nonlinear self-interaction of the scalar field as a higher-order effect compared to the non-minimal coupling of the scalar field and torsion. The tetrad field
\begin{eqnarray}
&&e^A_\mu=diag\left(1,a,a,a \right),
\label{e}
\end{eqnarray}
corresponds to isotropic and homogeneous Friedmann-Lemaitre-Robertson-Walker (FLRW) metric
\begin{eqnarray}
\label{FLRW}
&&ds^2=-dt^2+a^2\delta_{ij}dx^idx^j,
\end{eqnarray}
where $a=a(t)$ is the scale factor that dependents on cosmic time $t$ only.

For FLRW metric (\ref{e})--(\ref{FLRW}) expressions (\ref{T})--(\ref{TK}) imply following expression for torsion scalar
\begin{eqnarray}
\label{HT}
&&T=6H^{2}.
\end{eqnarray}

Background equations for action Eq.~(\ref{act1}) and tetrad field (\ref{e}) are \cite{Gonzalez-Espinoza:2019ajd,Gonzalez-Espinoza:2020azh,Gonzalez-Espinoza:2020jss}
\begin{eqnarray}
&&P-2XP_{,X}-3 H^2F=0,\\
\label{1}
&&P-3 H^2F-2\dot{H}F-2\dot{F}H=0,\\
\label{2}
&&P_{,\phi}-2XP_{,\phi X}+3H^2F_{,\phi}-3HP_{, X}\dot{\phi}\nonumber\\&&-\left[ P_{, X}+2P_{, XX}\right]\ddot{\phi}=0,
\label{3}
\end{eqnarray}
where an over dot denotes ordinary derivative concerning cosmic time $t$.

For brevity, we denote $P\approx P(\phi,X)$ and consider
\begin{eqnarray}
\label{P}
&&P=-\omega X+V(\phi)=-\frac{\omega}{2}\dot{\phi}^2+V(\phi),
\end{eqnarray}
where $\omega$ is a constant.
	
From Eqs.~(\ref{1})--(\ref{P}), we obtain
\begin{eqnarray}
\label{poten}
&&V(\phi)=3H^2F+\dot{H}F+\dot{F}H,\\
\label{pole}
&&\omega\dot{\phi}^2 =-2H\dot{F} - 2F\dot{H},\\
\label{field}
&&\omega\ddot{\phi}+3\omega H\dot{\phi}+V_{,\phi}+3H^2F_{,\phi}=0.
\end{eqnarray}

To note, for the case of minimal coupling ($F=1$), these models are reduced to the case of inflationary models based on TEGR.

We also note that of the three equations of cosmological dynamics (\ref{poten})--(\ref{field}) only two are independent. Thus, we can use any two equations in system (\ref{poten})--(\ref{field}) to analyze the models of cosmological inflation.

\section{Power-law parametrization of coupling function}\label{SEC3}

To estimate the influence of non-minimal coupling on the cosmological parameters, we consider some connection between coupling function $F=F(\phi)$ and Hubble parameter $H=H(t)$. We define the power-law connection as,
\begin{eqnarray}
\label{PLPM}
&&F(\phi(t))=\left(\frac{H(t)}{\lambda}\right)^{n},
\end{eqnarray}
where $\lambda$ is a positive constant and $n$ is the parameter that defines the influence of the non-minimal coupling between the scalar field and torsion.

For $n=0$ one has $F=1$, and action (\ref{act1}) is reduced to following form
\begin{eqnarray}
&&S\approx\int  d^4x e \left[\frac{1}{2}T+P(\phi,X) \right],
\label{actTEGR}
\end{eqnarray}
corresponding to the case of the teleparallel equivalent of general relativity~\cite{Gonzalez-Espinoza:2019ajd,Gonzalez-Espinoza:2020azh,Gonzalez-Espinoza:2020jss}.

The power-law connection between coupling function and Hubble parameter was considered earlier in~\cite{Fomin:2017sbt,Fomin:2020woj,Fomin:2022ozv} in the context of the model-independent approach to verification of the inflationary models based on the scalar-tensor gravity theories.
Also, relation (\ref{PLPM}) was considered in~\cite{Chervon:2023gio} for the case of the inflationary models based on the scalar-torsion gravity with arbitrary kinetic function $\omega=\omega(\phi)$. In~\cite{Chervon:2023gio} the constant $n$ characterized the influence of the non-minimal coupling of the scalar field and torsion on cosmological parameters, but it was not itself directly related to any cosmological parameter.
In the framework of the new approach, we will relate both free constant parameters in relationship (\ref{PLPM}) with the speed of propagation of scalar perturbations. This method of parameterizing the influence of the scalar-torsion modifications of TEGR leads to additional physical motivated constraints on the constants $n$ and $\lambda$ in relationship (\ref{PLPM}) compared to the approach proposed in~\cite{Chervon:2023gio}.
\color{black}

Also, we note, that relation (\ref{PLPM}) can be defined in parametric form for any solutions of dynamic equations (\ref{poten})--(\ref{pole}), and for the special case $n=0$, we obtain the minimal coupling for any Hubble parameter.

Now, Eqs.(\ref{poten})--(\ref{pole}) become
\begin{eqnarray}
\label{poten_3}
&&V(\phi)=3\lambda^{-n}H^{2+n}+\lambda^{-n}\dot{H}H^n(1+n),\\
\label{pole_3}
&&\omega\dot{\phi}^2=-2\lambda^{-n}\dot{H}H^n(n+1),
\end{eqnarray}
where $\omega=constant$ corresponding to the canonical (for $\omega>0$) and phantom (for $\omega<0$) scalar fields.

Thus, Eqs. (\ref{poten_3})--(\ref{pole_3}) can be rewritten as,
\begin{eqnarray}
\label{POT}
&&V(\phi)=\frac{H^{n}}{\lambda^{n}}\left(3H^2+\dot{H}(1+n)\right)\nonumber\\&&=
F(\phi)\left(3H^2+\dot{H}(1+n)\right),\\
\label{FIELD}
&&\omega\dot{\phi}^2=-2\left(\frac{H}{\lambda}\right)^{n}\dot{H}(n+1)
=-\frac{2}{\lambda^{n}}\frac{d}{dt}\left(H^{n+1}\right),
\end{eqnarray}
where
\begin{eqnarray}
\label{wlambda}
&&\omega=\lambda^{-n}=constant.
\end{eqnarray}
The convenient normalization of the constant $\omega$ implying canonical scalar field $\omega=1$ for the minimal coupling $n=0$. To note in minimal coupling, dynamic equations (\ref{POT})--(\ref{FIELD}) are reduced to the well-known expressions~\cite{Chervon:2019sey}
\begin{eqnarray}
\label{POT1}
&&\tilde{V}(\phi)=3h^2+\dot{h}=3h^2-\frac{1}{2}\dot{\phi}^{2},\\
\label{FIELD1}
&&\dot{\phi}^2=-2\dot{h},
\end{eqnarray}
where $h=h(t)>0$ is the Hubble parameter corresponding to the minimal coupling.

The quasi de-Sitter expansion of the early Universe $h\approx constant$ implies the following slow-roll condition $X=\frac{1}{2}\dot{\phi}^{2}\ll V$ at the inflationary stage.

We introduce the connection between $H$ and $h$ as,
\begin{eqnarray}
\label{HREL}
&&H=h^{\frac{1}{1+n}},\\
\label{HREL2}
&&\dot{H}=\left(\frac{1}{1+n}\right)\dot{h}h^{-\frac{n}{1+n}},
\end{eqnarray}	
Eqs. (\ref{POT})--(\ref{FIELD1}) can now be rewritten as,
\begin{eqnarray}
\label{POT2}
&&V(\phi)=\lambda^{-n}\left[3h^{\frac{2+n}{1+n}}+\dot{h}\right]=
\lambda^{-n}\left[3h^{\frac{2+n}{1+n}}-\frac{1}{2}\dot{\phi}^{2}\right],\\
\label{FIELD2}
&&\dot{\phi}^2=-2\dot{h}.
\end{eqnarray}

Therefore, the kinetic energy of the scalar field for non-minimal and minimal coupling is the same $X=\tilde{X}=\frac{1}{2}\dot{\phi}^{2}$ for relation (\ref{HREL}). This relation allows us to consider the fulfillment of the inflationary slow-roll conditions $\tilde{X}\ll \tilde{V}$ and $X\ll V$ for these two cases at once.
Also, relation (\ref{HREL}) implies the same evolution of the scalar field for both these cases $\phi(t)=\tilde{\phi}(t)$ up to the choice of a constant.

Further, using Ivanov-Salopek-Bond approach~\cite{Chervon:2019sey} based on the relation $\dot{h}=\dot{\phi}h_{,\phi}=\dot{\phi}\frac{dh(\phi)}{d\phi}$, one can rewrite Eqs. (\ref{POT2})--(\ref{FIELD2}) as
\begin{eqnarray}
\label{POT3}
&&V(\phi)=\lambda^{-n}\left\{3\left[h(\phi)\right]^{\frac{2+n}{1+n}}-2(h_{,\phi})^{2}\right\},\\
\label{FIELD3}
&&\dot{\phi}=-2h_{,\phi}.
\end{eqnarray}

Under condition $1+n>0$ for $\dot{h}<0$ one can have $\dot{H}<0$ as well. So, under condition $1+n>0$ the inflationary stage is being implemented for the case of non-minimal coupling if inflation occurs for the case of minimal coupling. The condition of inflationary stage $1+n>0$ restricts the possible values of the characterization constant as,
\begin{eqnarray}
\label{PC}
&&-1<n.
\end{eqnarray}

The coupling function (\ref{PLPM}) in terms of the Hubble parameter for minimal coupling is
\begin{eqnarray}
\label{COUPL}
&&F(\phi)=\lambda^{-n}\left[h(\phi)\right]^{\frac{n}{1+n}}.
\end{eqnarray}

Under the slow-roll conditions $\tilde{X}\ll \tilde{V}$ and $X\ll V$, from expressions (\ref{POT}), (\ref{POT1}) and (\ref{POT3}) we get
\begin{eqnarray}
\label{POTFF1}
&&\tilde{V}(\phi)\simeq3h^2,\\
\label{POTFF2}
&&V(\phi)\simeq3\lambda^{-n}H^{2+n}=3\lambda^{-n}h^{\frac{2+n}{1+n}}.
\end{eqnarray}

Thus, from Eqs. (\ref{POTFF1})--(\ref{POTFF2}) we obtain the following relation between potentials for non-minimal and minimal coupling
\begin{eqnarray}
\label{POTNEW}
&&V(\phi)\simeq3\lambda^{-n}\left(\frac{\tilde{V}(\phi)}{3}\right)^{\frac{2+n}{2(1+n)}}.
\end{eqnarray}

Also, from Eqs. (\ref{COUPL}) and (\ref{POTFF1}), we can obtain the relation between the coupling function and the potential for minimal coupling
\begin{eqnarray}
\label{FNEW}
&&F(\phi)\simeq\lambda^{-n}\left(\frac{\tilde{V}(\phi)}{3}\right)^{\frac{n}{2(1+n)}}.
\end{eqnarray}

Further, based on relations (\ref{HREL})--(\ref{HREL2}), we will express the slow-roll parameters for the case of non-minimal coupling in terms of the Hubble parameter $h=h(t)$ as,
\begin{eqnarray}
\label{EMIN}
&&\epsilon=-\frac{\dot{H}}{H^{2}}=-\frac{\dot{h}}{(1+n)}h^{-\frac{2+n}{1+n}},\\
\label{DMIN}
&&\delta=-\frac{\ddot{H}}{2H\dot{H}}=\frac{n\dot{h}}{2(1+n)}h^{-\frac{2+n}{1+n}}-\frac{\ddot{h}}{2\dot{h}}h^{-\frac{1}{1+n}}.
\end{eqnarray}

For the case $n=0$, from Eqs. (\ref{EMIN})--(\ref{DMIN}) one can have the slow-roll parameters for inflation with minimal coupling~\cite{Baumann:2014nda,Chervon:2019sey}
\begin{eqnarray}
\label{EMIN1}
&&\tilde{\epsilon}=-\frac{\dot{h}}{h^{2}}=2\left(\frac{h_{,\phi}}{h}\right)^{2},\\
\label{DMIN1}
&&\tilde{\delta}=-\frac{\ddot{h}}{2\dot{h}h}=2\frac{h_{,\phi\phi}}{h}.
\end{eqnarray}

Thus, to estimate the corrections induced by the non-minimal coupling between scalar field and torsion, it is necessary to determine two constant parameters: $\lambda$ and $n$.
We will determine the constant $\lambda$ based on the conditions for the completion of the inflationary stage.
Also, we will determine the lower limit on the parametrization constant $n$ based on the analysis of the cosmological perturbations parameters.

\section{Equation of state (EoS) parameter at the end of the inflationary stage}\label{SEC4}

Now, we shall determine the constant parameter $\lambda$ from the conditions on the EoS parameter at the end of the stage of cosmological inflation. In cosmological models with minimal coupling, the EoS parameter of a scalar field can be given as~\cite{Baumann:2014nda, Chervon:2019sey},
\begin{eqnarray}
\label{SP1}
&&\tilde{w}=\frac{\tilde{p}}{\tilde{\rho}}=\frac{\tilde{X}-\tilde{V}}{\tilde{X}+\tilde{V}}=-1-\frac{2}{3}\frac{\dot{h}}{h^{2}}=-1+\frac{2}{3}\tilde{\epsilon}.
\end{eqnarray}
where $\tilde{p}$ and $\tilde{\rho}$ are respectively the pressure and energy density of a scalar field.

At the inflationary stage, since $\tilde{\epsilon}\ll1$ one can interpret from Eq. (\ref{SP1}), $\tilde{w}\simeq-1$.
The end of the accelerated expansion stage of the early Universe is defined by condition $\tilde{\epsilon}=1$ with corresponding EoS parameter $\tilde{w}_{E}=-1/3$ which follows from (\ref{SP1}) for the case of the minimal coupling~\cite{Baumann:2014nda, Chervon:2019sey}.

For the case of non-minimal coupling ($n\neq 0$), the EoS parameter is
\begin{eqnarray}
\label{SP}
&&w=\frac{p}{\rho}=\frac{X-V}{X+V}=\frac{-3\lambda^{-n} h^{\frac{2+n}{1+n}}-\dot{h}(1+\lambda^{-n})}
{3\lambda^{-n} h^{\frac{2+n}{1+n}}-\dot{h}(1-\lambda^{-n})}.
\end{eqnarray}

Using Eq. (\ref{EMIN}) in Eq. (\ref{SP}), one can get
\begin{eqnarray}
\label{SP1A}
&&w=\frac{-3+(\lambda^{n}+1)(1+n)\epsilon}{3+(\lambda^{n}-1)(1+n)\epsilon}.
\end{eqnarray}

At the inflationary stage, since $\tilde{\epsilon}\ll1$ and $\epsilon\ll1$ from expressions (\ref{SP1}) and (\ref{SP1A}), one can find $\tilde{w}\simeq-1$ and $w\simeq-1$.

For the special case $n=2$, due to Eq. (\ref{SP1A}), at the end of inflation $\epsilon=1$ the EoS parameter is $w_{E}=1$ for any value of the constant parameter $\lambda$. This case corresponds to the post-inflationary stage of kinetic energy domination (kination), which precedes the further stage of radiation domination~\cite{Giovannini:2019oii,Tanin:2020qjw,Odintsov:2022sdk}.

For the case $n\neq2$, at the end of inflation $\epsilon=1$ from condition $w_{E}=-1/3$ and Eq. (\ref{SP1A}), we obtain following relation
\begin{eqnarray}
\label{WEND}
&&\lambda^{n}=\frac{2-n}{2(1+n)},
\end{eqnarray}
corresponding to the standard completion of the inflationary stage.

Again, from condition $\lambda^{n}>0$, we have the restrictions on the parametrization constant $-1<n<2$. Now, from Eqs. (\ref{POTNEW})--(\ref{FNEW}) and Eq. (\ref{WEND}), we obtain the following relations
\begin{eqnarray}
\label{POTNEW2}
&&V(\phi)\simeq\frac{6(1+n)}{(2-n)}\left(\frac{\tilde{V}(\phi)}{3}\right)^{\frac{2+n}{2(1+n)}},\\
\label{FNEW2}
&&F(\phi)\simeq\frac{2(1+n)}{(2-n)}\left(\frac{\tilde{V}(\phi)}{3}\right)^{\frac{n}{2(1+n)}},
\end{eqnarray}
which characterize the influence of the non-minimal coupling on the potential and define the coupling function.

To determine what type of completion of the inflationary stage is implemented for the models under consideration, it is necessary to determine the upper limit on the parametrization constant $n$ in addition to the lower limit (\ref{PC}). This estimate can be obtained based on the analysis of the influence of non-minimal coupling between a scalar field and torsion on the parameters of the cosmological perturbations in the inflationary models.
\section{Cosmological perturbations}\label{SEC5}

From the standpoint of inflationary paradigm, the origin of the large-scale structure of the Universe is due to quantum fluctuations of the scalar field and corresponding scalar perturbations of the space-time metric at the inflationary stage. Also, the theory of cosmological perturbations predicts the presence of relict gravitational waves or tensor perturbations in the present era which were also induced at the inflationary stage~\cite{Baumann:2014nda,Chervon:2019sey}.

The influence of cosmological perturbations on anisotropy and polarization of CMB makes it possible to determine the correctness of inflationary models based on observational constraints on the values of the parameters of cosmological perturbations~\cite{Baumann:2014nda,Chervon:2019sey}. Current observational data \cite{Planck:2018vyg,Tristram:2021tvh} suggest the constraints on the cosmological perturbation parameters as,
\begin{eqnarray}
\label{PARCONSTRAINTPS}
&&P_{S}=2.1\times10^{-9},\\
\label{PARCONSTRAINTNS}
&&n_{S}=0.9663\pm 0.0041,\\
\label{PARCONSTRAINTR}
&&r<0.032,
\end{eqnarray}
where $P_{S}$ represents the power spectrum of scalar perturbations, $n_{S}$ and $r$ respectively be the spectral index of scalar perturbations and tensor-to-scalar ratio.

It is obvious that the non-minimal coupling of the scalar field and torsion affects the evolution of cosmological perturbations and induces some corrections to the values of the parameters of cosmological perturbations relative to the case of minimal coupling.
The analysis of the evolution of cosmological perturbations in inflationary models based on TEGR and its modifications was studied earlier in Refs.~\cite{Chen:2010va,Golovnev:2018wbh,Raatikainen:2019qey,Gonzalez-Espinoza:2019ajd,Gonzalez-Espinoza:2020azh}.
In generalized scalar-torsion gravity, to define the perturbed metric, the cosmological perturbations for the inflationary models were studied based on Arnowitt-Deser-Misner (ADM) decomposition of the tetrad field~\cite{Gonzalez-Espinoza:2019ajd}.
The parameters of cosmological perturbations at the crossing of the Hubble radius $k=aH$ can be written as,
\begin{eqnarray}
\label{P1}
&&{\mathcal P}_{S}=\frac{H^2}{8\pi^2 Q_S c_{S}^3} = \frac{H^2}{8 \pi^2 F c_{S} \epsilon_s},\\
\label{P2}
&&{\mathcal P}_{T}=\frac{H^2}{2 \pi^2 Q_T}=\frac{2 H^2}{\pi^2 F},\\
\label{P3}
&&n_{S}-1=-2 \epsilon-\delta_{F}-\eta_{s}-s,\\
\label{P4}
&&n_{T}=-2 \epsilon-\delta_{F},\\
\label{P5}
&&r=\frac{{\mathcal P}_{T}}{{\mathcal P}_{S}}=16 c_s \epsilon_s,
\end{eqnarray}
where the stability parameters are
\begin{eqnarray}
\label{P6}
&&Q_S=\frac{w_1}{3 w_2^2 } \left( 9 w_2^2 + 4 w_1 w_3 \right),~~~~Q_{T}= \frac{1}{4} F,
\end{eqnarray}
and the slow-roll parameters are
\begin{eqnarray}
\label{P7}
&&\epsilon=-\frac{\dot{H}}{H^{2}},~~~~\delta=-\frac{\ddot{H}}{2H\dot{H}},~~~~\delta_{F}=\frac{\dot{F}}{H F},\\
\label{P7a}
&&\epsilon_{s}=\frac{Q_S c_{S}^2}{F},~~~~\eta_{s}=\frac{\dot{\epsilon}_{s}}{H\epsilon_{s}},~~~~s=\frac{\dot{c}_{S}}{H c_{S}},
\end{eqnarray}
The velocity of the scalar perturbations becomes,
\begin{eqnarray}
\label{P8}
&&c_S^2=\frac{3 \left(2 H w_1 w_2+ 4 \dot{w_1}w_2 -2 w_1 \dot{w_2}-w_2^2 \right)}{9 w_2^2 + 4 w_1 w_3}.
\end{eqnarray}

Here the slow-roll parameters satisfy
$\epsilon\ll1, \delta\ll1, \delta_{F}\ll1, \epsilon_{s}\ll1, \eta_{s}\ll1, s\ll1 $ during inflation. Also for $\dot{H}<0$, we can get $\epsilon>0$.

In the inflationary models for action (\ref{act}), the functions $w_{1}$, $w_{2}$, $w_{3}$ can be defined as~\cite{Gonzalez-Espinoza:2019ajd},
\begin{eqnarray}
\label{P9}
&&w_1=F,~~~~w_2=2HF,\\
\label{P10}
&&w_3=-3XP_{,X}-9H^2F=\frac{3}{2}\omega\dot{\phi}^{2}-9H^2F\nonumber\\&&=-3F\dot{H}-3H\dot{F}-9H^2F,
\end{eqnarray}
and the velocity of the tensor perturbations is equal to the speed of light in vacuum $c_{T}^2=1$.

We also note that in the case of scalar-torsion gravity violation of local Lorentz symmetry induces additional gauge degrees of freedom, which arises as a result of the different choice of the particular perturbed tetrad frame. These gauge degrees of freedom correspond to the additional scalar mode, the transverse vector mode, and the (pseudo) vector mode~\cite{Gonzalez-Espinoza:2019ajd,Gonzalez-Espinoza:2020azh}.
This property of cosmological inflation models based on scalar-torsion gravity was discussed earlier, for example, in~\cite{Chen:2010va, Golovnev:2018wbh, Raatikainen:2019qey,Gonzalez-Espinoza:2019ajd,Gonzalez-Espinoza:2020azh}.

However, in~\cite{Gonzalez-Espinoza:2019ajd} it was shown that for the inflationary models based on the action (\ref{act1}) Lorentz violating stage with essential contribution to inflation occurs deep inside the horizon, and these additional modes quickly decay with expansion of the early Universe.
Also, in this case, the power spectrum of the scalar perturbations is calculated as the two-point correlation function based on the standard second-order action implying negligible non-Gaussianity~\cite{Gonzalez-Espinoza:2019ajd}.
Thus, in accordance with~\cite{Gonzalez-Espinoza:2019ajd} we will consider standard scalar and tensor modes in the linear order of the cosmological perturbations theory at the crossing of the Hubble radius $(k=aH)$ for inflationary models under consideration.

\subsection{Velocities of scalar and tensor perturbations}\label{SEC5.1}

In general case, the velocity of propagation of scalar perturbations is a function of cosmic time [$c_S=c_S(t)$]. So, the corrections induced by the non-minimal coupling of scalar field and torsion must be considered for each specific model separately. Nevertheless, for the power-law connection between the non-minimal coupling function and the Hubble parameter $F=(H/\lambda)^{n}$ from Eqs. (\ref{P8})--(\ref{P10}), we obtain the following expression for the velocity of propagation of scalar perturbations
\begin{eqnarray}
\label{CS}
c_S^2=\frac{1-n}{1+n}=constant,
\end{eqnarray}
for an arbitrary parameter of the model.

For $c_S^2>0$, Eq. (\ref{CS}) leads to the following constraint on the constant,
\begin{eqnarray}
\label{CONSTRAINTS2}
-1<n<1,
\end{eqnarray}
which correspond to lower limit (\ref{PC}).

Thus, we have three possible regimes of scalar perturbations propagation, namely
\begin{enumerate}
  \item sub-luminal regime with $0<c_S<1$ for $0<n<1$,
  \item luminal regime with $c_S=1$ for $n=0$,
  \item super-luminal regime with $c_S>1$ for $-1<n<0$.
\end{enumerate}

\begin{figure}[ht]
	\centering
		\includegraphics[width=0.45\textwidth]{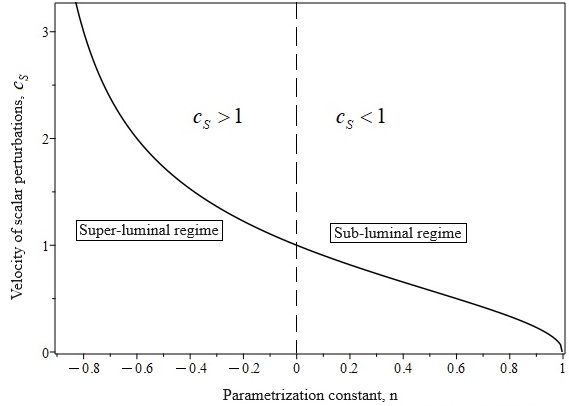}
	\caption{The dependence of the velocity of scalar perturbations from the value of the parametrization constant $n$.}
	\label{FIG1}
\end{figure}

Fig.\ref{FIG1} shows the dependence of the propagation speed of scalar perturbations on the parametrization constant $c_{S}=c_{S}(n)$. This dependence demonstrates that, in the general case, the models of cosmological inflation under consideration imply three different regimes of propagation of scalar perturbations: sub-luminal regime, luminal regime and super-luminal regime.

A super-luminal propagation cosmological perturbations is possible in the presence of the time-dependent homogeneous
scalar field~\cite{Mukhanov:2005sc,Mukhanov:2005bu,Babichev:2007dw} and for different modified gravity theories~\cite{DeFelice:2009ak,Koh:2016abf,Mironov:2020pqh} as well. Also, we note that due to constraint (\ref{CONSTRAINTS2}), the possible post-inflationary kination stage (for $n=2$) can not be realized in these cosmological models.

The difference in propagation velocities of scalar and tensor perturbations leads to the fact that the Hubble radius was crossed by them in different wavelengths, namely
\begin{eqnarray}
\label{WL1}
&&c_{S}k_{S}=aH,~~~~k_{S}=\frac{2\pi a}{\lambda_{S}},\\
\label{WL2}
&&c_{T}k_{T}=aH,~~~k_{T}=\frac{2\pi a}{\lambda_{T}},\\
\label{WL3}
&&\frac{c_{S}}{c_{T}}=\frac{k_{T}}{k_{S}}=\frac{\lambda_{S}}{\lambda_{T}}=\sqrt{\frac{1-n}{1+n}},
\end{eqnarray}
Whereas, in case of the minimal coupling for which $c_{S}=c_{T}=1$, the wavelengths of scalar and tensor perturbations are equal ($\lambda_{S}=\lambda_{T}$).

On the basis of expressions (\ref{WEND}) and (\ref{CS}) we can define constant model's parameters as follows
\begin{eqnarray}
\label{WL4}
&&n=\frac{1-c_S^2}{1+c_S^2},\\
\label{WL5}
&&\lambda^{n}=\frac{1}{4}\left(1+3c_S^2\right).
\end{eqnarray}

For the case of the minimal coupling with velocity of the scalar perturbations $c_S^2=1$ one has $n=0$ and $\lambda=1$.

Therefore, initial power-law connection (\ref{PLPM}) between coupling function and the Hubble parameter can be redefined in terms of the velocity of propagation of scalar perturbations as
\begin{eqnarray}
\label{PLPMCS}
F(\phi(t))=\left(\frac{4}{1+3c_S^2}\right)\left(H(t)\right)^{\frac{1-c_S^2}{1+c_S^2}}.
\end{eqnarray}

Thus, the influence of non-minimal coupling between scalar field and torsion on the cosmological parameters can be parameterized by the speed of the scalar perturbations.
However, for convenience of presenting the results we will use parameters $n$ and $\lambda$ taking into account relations (\ref{WL4})--(\ref{WL5}).

\subsection{Parameters of cosmological perturbations}\label{SEC5.2}

For the case $F=(H/\lambda)^{n}$, from Eq. (\ref{P6}), we obtain
\begin{eqnarray}
\label{QS}
&&Q_S=-(1+n)\frac{\dot{H}}{H^{2}}\left(\frac{H}{\lambda}\right)^{n}\nonumber\\&&=(1+n)\left(\frac{H}{\lambda}\right)^{n}\epsilon>0,\\
\label{QT}
&&Q_{T}=\frac{1}{4}F=\frac{1}{4}\left(\frac{H}{\lambda}\right)^{n}>0.
\end{eqnarray}

The conditions $Q_S>0$ and $Q_{T}>0$ indicate that there are no Laplacian instabilities and ghosts in these models~\cite{Gonzalez-Espinoza:2019ajd}.
Also, the stability of scalar perturbations is defined by condition $c^{2}_{S}>0$ corresponding to constraint (\ref{CONSTRAINTS2}).

From (\ref{CS}) and (\ref{QS}) we get
\begin{eqnarray}
\label{QSCS}
&&Q_S=\left(\frac{2}{1+c^{2}_{S}}\right)\left(\frac{H}{\lambda}\right)^{n}\epsilon>0.
\end{eqnarray}

Thus, in this case, scalar perturbations are stable in the super-luminal regime as well. However, it will be shown below that the super-luminal regime of propagation of scalar perturbations cannot be realized in these models.

For $F=(H/\lambda)^{n}$, the slow-roll parameters Eqs. (\ref{P7})--(\ref{P7a}) can be related as,
\begin{eqnarray}
\label{SRPAR}
&&\epsilon_{s}=(1-n)\epsilon,~~~~~~\delta_{F}=-n\epsilon,\nonumber\\&&\eta_{s}=-2\delta+2\epsilon,~~~~~~s=0.
\end{eqnarray}

Thus, the cosmological perturbations (\ref{P1})--(\ref{P5}) parameters can be written as
\begin{eqnarray}
\label{PERTG2}
&&{\mathcal P}_{S}=\frac{(2-n)H^{2-n}_{\ast}}{8\pi^{2}(1-n^{2})\sqrt{\frac{1-n}{1+n}}\,\epsilon_{\ast}},\\
\label{PERTG2A}
&&{\mathcal P}_{T}=\frac{2\lambda^{n}}{\pi^{2}}H^{2-n}_{\ast},\\
\label{PERTG3}
&&n_{S}-1=(n-4)\epsilon_{\ast}+2\delta_{\ast},\\
\label{PERTG4}
&&r=16(1-n)\sqrt{\frac{1-n}{1+n}}\,\epsilon_{\ast},\\
\label{PERTG3A}
&&n_{T}=-(2-n)\epsilon_{\ast}.
\end{eqnarray}

For the case of the minimal coupling, we obtain
\begin{eqnarray}
\label{PERTG2mc}
&&\tilde{{\mathcal P}}_{S}=\frac{h^{2}_{\ast}}{8\pi^{2}\,\tilde{\epsilon}_{\ast}},\\
\label{PERTG2Amc}
&&\tilde{{\mathcal P}}_{T}=\frac{2h^{2}_{\ast}}{\pi^{2}},\\
\label{PERTG3mc}
&&\tilde{n}_{S}-1=-4\tilde{\epsilon}_{\ast}+2\tilde{\delta}_{\ast},\\
\label{PERTG4mc}
&&\tilde{r}=16\tilde{\epsilon}_{\ast},\\
\label{PERTG3Amc}
&&\tilde{n}_{T}=-2\tilde{\epsilon}_{\ast}.
\end{eqnarray}

All parameters of cosmological perturbations are considered at the crossing of the Hubble radius ($k=aH$ and $\tilde{k}=\tilde{a}h$ for some cosmic times $\tilde{t}=\tilde{t}_{\ast}$ and $t=t_{\ast}$, where $t_{\ast}\neq\tilde{t}_{\ast}$ and, correspondingly, $H^{n+1}_{\ast}\neq h_{\ast}$).

Since both minimal and non-minimal coupling cosmological models satisfy condition (\ref{PARCONSTRAINTPS}), the relation between the cosmic times of the crossing of the Hubble radius $\xi=\left(t_{\ast}/\tilde{t}_{\ast}\right)$ can be obtained from the equation for certain inflationary models
\begin{eqnarray}
\label{ConnectionTimes}
&&{\mathcal P}_{S}=\tilde{{\mathcal P}}_{S}=2.1\times10^{-9},
\end{eqnarray}
where $\xi=1$ for the case $n=0$.

Since condition (\ref{PARCONSTRAINTPS}) can always be satisfied by choosing the time of crossing the Hubble radius or model's parameters, the spectral index of scalar perturbations and the tensor-scalar ratio as the main parameters necessary for verifying inflationary models by observational constraints (\ref{PARCONSTRAINTNS})--(\ref{PARCONSTRAINTR}).
Also, in accordance with restrictions (\ref{CONSTRAINTS2}), condition $\epsilon_{\ast}>0$ and expression (\ref{PERTG3A}), on has red tilt tensor spectrum $n_{T}<0$ for arbitrary cosmological models under consideration.

\subsection{Transition to inflation based on $f(T)$-gravity}\label{SEC5.3}

Now let us consider the possibility of reducing the model based on action (\ref{act}) to the case of $f(T)$-gravity.
This consideration is motivated by the fact that $f(T)$-gravity model includes the super-luminal mode that
could arise from one of the extra degrees of freedom as shown in~\cite{Ong:2013qja,Izumi:2012qj}.
Thus, we will testify the possibility to obtain the additional gauge scalar modes by transition to $f(T)$-gravity in the inflationary models under consideration.

Under the condition of zero kinetic energy of a scalar field $\omega=0$ and for relations
\begin{eqnarray}
\label{ACTION1A}
&&F(\phi)=\phi=\frac{df(T)}{dT},\\
\label{ACTION2B}
&& V(\phi)=f(T(\phi))-\frac{\phi}{2}T(\phi),
\end{eqnarray}
action (\ref{act1}) is reduced to the following form
\begin{eqnarray}
\label{ACTION3C}
&&S=\int  d^4x e f(T),
\end{eqnarray}
corresponding to effective $f(T)$-gravity model.

Also, using expressions (\ref{act1}), (\ref{HT}) and (\ref{PLPM}) under conditions $\omega=0$ and $V=0$ we can obtain partial case of $f(T)$-gravity model, namely
\begin{eqnarray}
\label{ACTION4D}
&&S\sim\int  d^4x e T^{n/2}.
\end{eqnarray}

However, from expressions (\ref{P6}) and (\ref{P9})--(\ref{P10}) for $\omega=0$ one has $Q_{S}=0$. This result corresponds to the absence of scalar perturbations. Also, from (\ref{QS}) and $Q_{S}=0$ we can conclude that condition (\ref{PLPM}) implies that cosmological dynamics for $f(T)$-gravity corresponds to the pure de Sitter stage with $H=constant$ in the models under consideration. As one can see from relation (\ref{HT}), for constant Hubble parameter $f(T)$-gravity degenerates to the case $f(T)=constant$ in FLRW space-time without scalar field.

Thus, to construct correct inflationary models with quasi de Sitter dynamics and non-zero scalar perturbations we will consider condition $\omega\neq0$.

\section{Condition of the exit from inflation}\label{SEC6}

Expressions for the slow-roll parameters (\ref{EMIN})--(\ref{DMIN1}) leads to the following relation
\begin{eqnarray}
\label{ED2}
&&\frac{\delta}{\epsilon}=(1+n)\frac{\tilde{\delta}}{\tilde{\epsilon}}-\frac{n}{2}.
\end{eqnarray}

Since relation (\ref{ED2}) must be satisfied for any cosmic time, we get the following conditions
\begin{eqnarray}
\label{CONDT1}
&&\tilde{\delta}/\tilde{\epsilon}=constant,\\
\label{CONDT2}
&&\delta/\epsilon=constant,
\end{eqnarray}
which restricts the possible types of inflationary models.

Based on the definitions of the slow-roll parameters (\ref{EMIN})--(\ref{DMIN}) and the relation (\ref{CONDT2}), we get following types of inflationary dynamics:

1. Intermediate inflation with following Hubble parameter and scale factor
\begin{eqnarray}
\label{INT1}
&&H(t)=A\gamma t^{\gamma-1},\\
\label{INT2}
&&a(t)=a_{s}\exp(At^{\gamma}),
\end{eqnarray}
where $a_{s}$ is the scale factor at the beginning of inflation, $A>0$ and $0<\gamma<1$.

Intermediate inflation was considered earlier in~\cite{Barrow:1990td,Barrow:1993zq,Herrera:2014mca,Rezazadeh:2015dza,Jamil:2013nca,Oikonomou:2017isf}.
Intermediate inflation based on $f(T)$-gravity was also considered in~\cite{Jamil:2013nca,Oikonomou:2017isf,Rezazadeh:2015dza}.

Slow-roll parameters (\ref{EMIN})--(\ref{DMIN}) corresponding to Hubble parameter (\ref{INT1}) are
\begin{eqnarray}
\label{INT3}
&&\epsilon=\frac{(1-\gamma)}{A\gamma t^{\gamma}},\\
\label{INT4}
&&\delta=\frac{(2-\gamma)}{2A\gamma t^{\gamma}}.
\end{eqnarray}

Since slow-roll parameter $\epsilon$ is a decreasing function, it is impossible to simultaneously fulfill the slow-roll condition $\epsilon\ll1$ during inflation and exit from inflation $\epsilon_{e}=\epsilon(t=t_{e})=1$. In this case,  $t_{e}$ is the time of the end of inflation, which is defined from condition $\ddot{a}=0$ or $\epsilon=1$, since
\begin{eqnarray}
\label{INT5}
&&\ddot{a}=aH^{2}(1-\epsilon).
\end{eqnarray}

Thus, there is no exit from inflation by violation of the slow-roll conditions for the case of intermediate inflation.

2. Power-law inflation with the following Hubble parameter and scale factor
\begin{eqnarray}
\label{INT1PL}
&&H(t)=\frac{m}{t},\\
\label{INT2PL}
&&a(t)=a_{s}t^{m},
\end{eqnarray}
where $m>1$.

Power-law inflation was considered earlier in large numbers of works (see, for example, in~\cite{Chervon:2019sey, Chervon:2017kgn}, for a review).
Also, power-law inflation based on $f(T)$-gravity was considered in~\cite{Rezazadeh:2015dza}.

Slow-roll parameters (\ref{EMIN})--(\ref{DMIN}) corresponding to Hubble parameter (\ref{INT1PL}) are
\begin{eqnarray}
\label{INT3PL}
&&\epsilon=\delta=\frac{1}{m}=constant.
\end{eqnarray}

Thus, in this case, there is no exit from inflation by violation of the slow-roll conditions as well.

To obtain the exit from inflation for the case of intermediate and power-law inflation the possibility of implementing the additional reheating process for modified potentials compared to the TEGR case was considered in~\cite{Rezazadeh:2015dza}.
However, we can consider the third possible type of inflationary dynamics, which implies the exit from inflation by violation of the slow-roll conditions.
This type of exit from inflation corresponds to the similar types of potentials as for the case of TEGR.

3. Inflation with linear deviation from pure de Sitter expansion and following Hubble parameter and scale factor
\begin{eqnarray}
\label{INT6}
&&H(t)=\lambda-\mu t,\\
\label{INT7}
&&a(t)=a_{s}\exp\left(\lambda t-\frac{\mu}{2}t^{2}\right),
\end{eqnarray}
where $\lambda>0$ and $\mu>0$ are and some constants.

Slow-roll parameters (\ref{EMIN})--(\ref{DMIN}) corresponding to Hubble parameter (\ref{INT6}) are
\begin{eqnarray}
\label{INT8}
&&\epsilon=\frac{\mu}{H^{2}}=\mu\left(\lambda-\mu t\right)^{-2},~~~~\delta=0.
\end{eqnarray}

Since slow-roll parameter (\ref{INT8}) is growing function of the cosmic time, in this case, the end of inflation by violation of the slow-roll conditions $\epsilon_{e}=1$ can be realised for $t_{e}=\frac{\lambda}{\mu}\pm\frac{1}{\sqrt{\mu}}$.

Corresponding condition (\ref{ED2}) for $\delta=0$ is
\begin{eqnarray}
\label{INTCONDITION2}
&&(1+n)\frac{\tilde{\delta}}{\tilde{\epsilon}}-\frac{n}{2}=0.
\end{eqnarray}

Thus, we will consider (\ref{INTCONDITION2}) as the condition of the exit from inflation in cosmological models under consideration.

\section{Inflationary models with linear dependence $r\sim(1-n_{S})$}\label{SEC7}

Now we will consider the dependence of the tensor-scalar ratio on the spectral index of scalar perturbations $r=r(1-n_{S})$. Since the value of the spectral index of scalar perturbations, $n_{S}\simeq0.97$ and $1-n_{S}\simeq0.03\ll1$, we can write the dependence $r=r(1-n_{S})$ as follows,
\begin{eqnarray}
\label{pert13}
&&r=\sum^{\infty}_{k=0}\beta_{k}(1-n_{S})^{k}=\beta_{0}+\beta_{1}(1-n_{S})+\nonumber\\&&\beta_{2}(1-n_{S})^{2}+...,
\end{eqnarray}
where $(1-n_{S})\ll1$ is the small parameter of expansion and $\beta_{k}$ are the constant coefficients.

In zeroth order of this expansion $r(0)=\beta_{0}=0$ we consider condition $r(n_{S}=1)=0$ for the flat Harrison-Zeldovich spectrum \cite{Harrison:1969fb,Zeldovich:1969sb}.
Thus, we can rewrite expression (\ref{pert13}) in the following form
\begin{eqnarray}
\label{RNSEXPANSION}
&&r=\sum^{\infty}_{k=1}\beta_{k}(1-n_{S})^{k}=\beta_{1}(1-n_{S})\nonumber\\&&+\beta_{2}(1-n_{S})^{2}+...
\end{eqnarray}

The first term in expression (\ref{RNSEXPANSION}) makes the main contribution to the value of the tensor-scalar ratio.
Thus, we can consider dependence
\begin{eqnarray}
\label{RNSEXPANSION2}
&&r\simeq\beta_{1}(1-n_{S}),
\end{eqnarray}
as the condition of the first order inflationary models with negligible high-order terms in (\ref{RNSEXPANSION}).

The inflationary models of the second-order with dependence
\begin{eqnarray}
\label{RNSEXPANSION3}
&&r\simeq\beta_{2}(1-n_{S})^{2},
\end{eqnarray}
were considered, for example, in~\cite{Fomin:2020woj,Fomin:2022ozv,Fomin:2020caa}.
In this case $\beta_{1}=0$, and terms above the second order are negligible ones.
\color{black}

Now, we consider the inflationary models with minimal coupling and constant relation between slow-roll parameters during inflation
\begin{eqnarray}
\label{FIRST1}
&&\frac{\tilde{\delta}}{\tilde{\epsilon}}=\frac{\tilde{\delta}_{\ast}}{\tilde{\epsilon}_{\ast}}=constant,
\end{eqnarray}
which follows from (\ref{CONDT1}).

After substituting relation (\ref{FIRST1}) into (\ref{PERTG3mc})--(\ref{PERTG4mc}), we obtain a linear relationship between the tensor-scalar ratio and the spectral index of scalar perturbations
\begin{eqnarray}
\label{FIRST2}
&&\tilde{r}=\tilde{\beta}_{1}(1-\tilde{n}_{S}).
\end{eqnarray}

Also, based on expressions (\ref{FIELD1}), (\ref{POTFF1}) and (\ref{EMIN1})--(\ref{DMIN1}), we can express the slow-roll parameters in terms of potential in the slow-roll approximation~\cite{Chervon:2019sey}
\begin{eqnarray}
\label{FIRST4}
&&\tilde{\epsilon}=-\frac{\dot{h}}{h^{2}}=2\left(\frac{h_{,\phi}}{h}\right)^{2}\simeq\frac{1}{2}\left(\frac{\tilde{V}_{,\phi}}{\tilde{V}}\right)^{2},\\
\label{FIRST5}
&&\tilde{\delta}=-\frac{\ddot{h}}{2\dot{h}h}=2\frac{h_{,\phi\phi}}{h}\simeq\frac{\tilde{V}_{,\phi\phi}}{\tilde{V}}
-\frac{1}{2}\left(\frac{\tilde{V}_{,\phi}}{\tilde{V}}\right)^{2},
\end{eqnarray}
where we use the following relations
\begin{eqnarray}
\label{RELED}
&&\dot{h}=-2h^{2}_{,\phi},~~~~~\ddot{h}=8h^{2}_{,\phi}h_{,\phi\phi},~~~~~h\simeq\left(\frac{V}{3}\right)^{1/2}.
\end{eqnarray}

Thus, from condition (\ref{FIRST1}) and equations (\ref{FIRST4})--(\ref{FIRST5}), we obtain two types of inflationary potential:
\begin{eqnarray}
\label{POTF1}
&&\tilde{V}(\phi)\sim\phi^{p},~~~~~~~~~~~~~~~~~{\text for}~~~\tilde{\delta}/\tilde{\epsilon}\neq 1,\\
\label{POTF2}
&&\tilde{V}(\phi)\sim\exp(-\alpha\phi),~~~~{\text for}~~~\tilde{\delta}/\tilde{\epsilon}=1.
\end{eqnarray}
where $p$ and $\alpha$ are a some constants.

Models of cosmological inflation based on these potentials have been considered previously in a large number of papers (see, for example, in~\cite{Martin:2013tda,Martin:2013nzq,Gron:2018rtj,Mishra:2018dtg,Avdeev:2022ilo,Odintsov:2023weg,Odintsov:2023rqf}).

Also, from (\ref{POTFF2}), (\ref{EMIN})--(\ref{DMIN}) and (\ref{RELED}), we obtain the following expressions for the slow-roll parameters for the case of non-minimal coupling
\begin{eqnarray}
\label{POTF3}
&&\epsilon= \frac{2 \left(\frac{d}{d \phi}h \left(\phi \right)\right)^{2} h \left(\phi \right)^{-\frac{2+n}{1+n}}}{1+n},\\
\label{POTF4}
&&\delta= -\frac{2 n \left(\frac{d}{d \phi}h \left(\phi \right)\right)^{2} h \left(\phi \right)^{-\frac{2+n}{1+n}}}{2+2 n}\nonumber\\&&+2 \left(\frac{d^{2}}{d \phi^{2}}h\! \left(\phi \right)\right) h \left(\phi \right)^{-\frac{1}{1+n}},\\
\label{POTF5}
&&h\left(\phi\right)\simeq\left(\frac{\lambda^{n}}{3}V\left(\phi\right)\right)^{\frac{1+n}{2+n}}.
\end{eqnarray}

Using these expressions, we will check the compliance with condition $\delta/\epsilon=const$ in the slow-roll approximation, since relation (\ref{ED2}) was obtained for exact expressions of the slow-roll parameters.

\subsection{Inflationary models with exponential potential}\label{SEC7.1}

Now, we consider the inflationary model with exponential potential~\cite{Martin:2013tda,Martin:2013nzq,Gron:2018rtj}
\begin{eqnarray}
\label{E1}
&&\tilde{V}(\phi)=\tilde{V}_{0}\exp(-\alpha\phi),
\end{eqnarray}
where $\tilde{V}_{0}$ and $\alpha$ are a some positive constants.

From Eqs. (\ref{FIRST4})--(\ref{FIRST5}), we obtain
\begin{eqnarray}
\label{E2}
&&\tilde{\epsilon}=\tilde{\epsilon}_{\ast}=\frac{\alpha^{2}}{2}>0,\\
\label{E3}
&&\frac{\tilde{\delta}}{\tilde{\epsilon}}=\frac{\tilde{\delta}_{\ast}}{\tilde{\epsilon}_{\ast}}=1.
\end{eqnarray}

Since relation (\ref{E3}) doesn't contain additional constant parameters, we can immediately determine the dependence of the tensor-scalar ratio on the spectral index of scalar perturbations. This dependence will be valid for any e-fold number including $50<\tilde{N}<70$.

Thus, from Eqs. (\ref{PERTG3mc})--(\ref{PERTG4mc}) and Eq. (\ref{E3}) we obtain
\begin{eqnarray}
\label{E4}
&&\tilde{r}=8(1-\tilde{n}_{S}).
\end{eqnarray}

After substituting $\tilde{n}_{S}\simeq0.97$ into Eq. (\ref{E4}) one can have $\tilde{r}\simeq0.24$.
Thus, the inflationary model with exponential potential (\ref{E1}) doesn't satisfy the observational constraint Eq. (\ref{PARCONSTRAINTR}) for the case of minimal coupling.

For the case of non-minimal coupling, from equations Eqs. (\ref{POTNEW2})--(\ref{FNEW2}) and Eq. (\ref{E1}) we get
\begin{eqnarray}
\label{Q7}
&&V(\phi)\simeq V_{0}\exp\left(-\frac{\alpha(2+n)}{2(1+n)}\phi\right),\\
\label{Q8}
&&F(\phi)\simeq F_{0}\exp\left(-\frac{n\phi}{2(1+n)}\right),
\end{eqnarray}
where
\begin{eqnarray}
\label{Q9}
&&V_{0}=\frac{6(1+n)}{(2-n)}\left(\frac{\tilde{V}_{0}}{3}\right)^{\frac{2+n}{2(1+n)}},\\
\label{Q10}
&&F_{0}=\frac{2(1+n)}{(2-n)}\left(\frac{\tilde{V}_{0}}{3}\right)^{\frac{2+n}{2(1+n)}}.
\end{eqnarray}

After substituting (\ref{Q7}) into (\ref{POTF3})--(\ref{POTF5}) we obtain
\begin{eqnarray}
\label{E5}
&&\frac{\delta}{\epsilon}=\frac{\delta_{\ast}}{\epsilon_{\ast}}=1+\frac{n}{2}=const,
\end{eqnarray}
that coincides with relation (\ref{ED2}).

Further, from (\ref{PERTG3})--(\ref{PERTG4}) and (\ref{E5}) we obtain
\begin{eqnarray}
\label{E6}
&&r=8\sqrt{\frac{1-n}{1+n}}\,(1-n_{S}).
\end{eqnarray}

For $n_{S}\simeq0.97$ and $r<0.032$, from (\ref{E6}) we get $0<n<0.965$.
Also, for $0<n<0.965$ corresponding velocity of the scalar perturbations (\ref{CS}) takes the values $0<c_{S}<0.13$.

On the other hand, from conditions (\ref{INTCONDITION2}) and (\ref{E3}) one has $n=-2$. Thus, this model does not imply an exit from inflation and cannot be considered as a relevant one.

\subsection{Inflationary models with power-law potential and minimal coupling}\label{SEC7.2}

Now, we consider the inflationary model with power-law potential~\cite{Martin:2013tda,Martin:2013nzq,Gron:2018rtj,Mishra:2018dtg,Avdeev:2022ilo,Odintsov:2023weg,Odintsov:2023rqf}
\begin{eqnarray}
\label{GPL}
&&\tilde{V}(\phi)=\tilde{V}_{0}\phi^{p},
\end{eqnarray}
where $\tilde{V}_{0}$ and $p$ are a some constants.

From (\ref{FIRST4})--(\ref{FIRST5}), we obtain
\begin{eqnarray}
\label{GPL1}
&&\tilde{\epsilon}=\frac{p^{2}}{2\phi^{2}}>0,\\
\label{GPL2}
&&\frac{\tilde{\delta}}{\tilde{\epsilon}}=\frac{\tilde{\delta}_{\ast}}{\tilde{\epsilon}_{\ast}}=\left(\frac{p-2}{p}\right)=const.
\end{eqnarray}

Since the slow-roll parameters (\ref{GPL1})--(\ref{GPL2}) contain an unknown arbitrary constant $p$, in this case, it is necessary to obtain an additional dependence of the slow-roll parameters on the e-folds number $\tilde{N}$.

Firstly, we define the value of the scalar field at the end of inflation $\phi=\phi_{e}$.
From condition $\epsilon=1$ and expression (\ref{GPL1}) we obtain
\begin{eqnarray}
\label{PLI1}
&&\phi^{2}_{e}=\frac{p^{2}}{2}.
\end{eqnarray}

Secondly, we define the e-folds number between the crossing of the Hubble radius and the end of inflation~\cite{Baumann:2014nda,Chervon:2019sey}
\begin{eqnarray}
\label{PLI2}
&&\tilde{N}=\int^{\tilde{t}_{e}}_{\tilde{t}_{\ast}} h(t)dt=\int^{\phi_{e}}_{\phi_{\ast}} \frac{h}{\dot{\phi}}d\phi=
-\frac{1}{2}\int^{\phi_{e}}_{\phi_{\ast}} \frac{h}{h_{,\phi}}d\phi \nonumber\\&&\simeq
-\int^{\phi_{e}}_{\phi_{\ast}} \frac{V}{V_{,\phi}}d\phi.
\end{eqnarray}

Further after substituting (\ref{GPL}) into (\ref{PLI2}), we obtain
\begin{eqnarray}
\label{PLI3}
&&\tilde{N}\simeq\frac{1}{2p}\left(\phi^{2}_{\ast}-\phi^{2}_{e}\right)=\frac{1}{2p}\left(\phi^{2}_{\ast}-\frac{p^{2}}{2}\right).
\end{eqnarray}

Thus from Eq. (\ref{PLI3}), at Hubble radius crossing, we obtain the value of the scalar field a
\begin{eqnarray}
\label{PLI4}
&&\phi^{2}_{\ast}=\frac{p}{2}\left(p+4\tilde{N}\right).
\end{eqnarray}

From Eqs. (\ref{GPL1})--(\ref{GPL2}) and Eq. (\ref{PLI4}) also, at the Hubble radius, the slow-roll parameters can be obtained as,
\begin{eqnarray}
\label{PLI5}
&&\tilde{\epsilon}_{\ast}=\frac{p}{p+4\tilde{N}},~~~~~
\tilde{\delta}_{\ast}=\frac{p-2}{p+4\tilde{N}}.
\end{eqnarray}

From Eqs. (\ref{PERTG3mc})--(\ref{PERTG4mc}) and Eq. (\ref{PLI5}) we obtain the following expressions for the spectral tilt of the scalar perturbations and tensor-to-scalar ratio
\begin{eqnarray}
\label{PLI5A}
&&1-\tilde{n}_{S}=\frac{2(p+2)}{p+4\tilde{N}},\\
\label{PLI5B}
&&\tilde{r}=\frac{16p}{p+4\tilde{N}}.
\end{eqnarray}

Thus, from equations Eqs. (\ref{PLI5A})--(\ref{PLI5B}) we obtain following dependencies of the tensor-to-scalar ratio from the spectral index of the scalar perturbations
\begin{eqnarray}
\label{GPL8}
&&\tilde{r}=\frac{8p}{p+2}(1-\tilde{n}_{S}),\\
\label{GPL9}
&&\tilde{r}=\frac{16\left[(1-\tilde{n}_{S})\tilde{N}-1\right]}{2\tilde{N}-1}.
\end{eqnarray}

Due to the observational value of the spectral tilt of the scalar perturbations $\tilde{n}_{S}=0.9663\pm 0.0041$ for a fairly wide range of the e-folds number $50<\tilde{N}<70$, from (\ref{GPL9}) we obtain $0.08<\tilde{r}<0.20$.
Thus, for the minimal coupling, these models don't correspond to the observational constraint (\ref{PARCONSTRAINTR}).

Also, from expression (\ref{PLI5A}), we obtain:
\begin{eqnarray}
\label{GPL10}
&&p_{min}(\tilde{n}_{S}=0.9704, \tilde{N}=50)\simeq1,\\
\label{GPL11}
&&p_{max}(\tilde{n}_{S}=0.9622, \tilde{N}=70)\simeq3.34.
\end{eqnarray}

Thus, in the slow-roll approximation, we can estimate the values of the parameter $p$ as $1<p<3.4$.

\subsection{Inflationary models with power-law potential and non-minimal coupling}\label{SEC7.3}

Now, we estimate the influence of non-minimal coupling between the scalar field and torsion on the cosmological parameters. From Eqs. (\ref{POTNEW2})--(\ref{FNEW2}) and Eq. (\ref{GPL}) we obtain the following expressions for the potential and non-minimal coupling function
\begin{eqnarray}
\label{POTNEW3}
&&V(\phi)\simeq V_{0}\phi^{k},\\
\label{FNEW3}
&&F(\phi)\simeq F_{0}\phi^{m},
\end{eqnarray}
where
\begin{eqnarray}
\label{CONST1}
&&V_{0}=\frac{6(1+n)}{(2-n)}\left(\frac{\tilde{V}_{0}}{3}\right)^{\frac{2+n}{2(1+n)}},~~~k=\frac{p(2+n)}{2(1+n)},\\
\label{CONST2}
&&F_{0}=\frac{2(1+n)}{(2-n)}\left(\frac{\tilde{V}_{0}}{3}\right)^{\frac{n}{2(1+n)}},~~~m=\frac{pn}{2(1+n)}.
\end{eqnarray}

After substituting Eq.(\ref{POTNEW3}) and Eq. (\ref{CONST1}) into Eqs. (\ref{POTF3})--(\ref{POTF5}) we obtain
\begin{eqnarray}
\label{CONST3}
&&\frac{\delta}{\epsilon}=\frac{\delta_{\ast}}{\epsilon_{\ast}}=
(1+n)\left(\frac{p-2}{p}\right)-\frac{n}{2}=const,
\end{eqnarray}
that coincides with relation (\ref{ED2}).

Also, from Eqs. (\ref{PERTG3})--(\ref{PERTG4}) and Eq. (\ref{CONST3}) we get following dependence tensor-to-scalar ratio from spectral tilt of the scalar perturbations
\begin{eqnarray}
\label{PLIN3}
&&r=\frac{8p\sqrt{\frac{1-n}{1+n}}(1-n)(1-n_{S})}{2(1+n)+p(1-n)}.
\end{eqnarray}

For $n=0$ this dependence is reduced to Eq. (\ref{GPL8}).

Taking into account condition of exit from inflation (\ref{INTCONDITION2}) and expression (\ref{GPL2}), we obtain
\begin{eqnarray}
\label{PLIN3A}
&&p=\frac{4(1+n)}{(2+n)}.
\end{eqnarray}

After substituting (\ref{PLIN3A}) into (\ref{POTNEW3})--(\ref{CONST2}) we obtain
\begin{eqnarray}
\label{POTNEW3B}
&&V(\phi)\simeq V_{0}\phi^{2},\\
\label{FNEW3B}
&&F(\phi)\simeq F_{0}\phi^{\frac{2n}{2+n}}.
\end{eqnarray}

Thus, under condition of exit from inflation one has quadratic potential (\ref{POTNEW3B}) for any value of the constant $n$, which defines the influence of the non-minimal coupling between the scalar field and torsion.

Also, from (\ref{PLIN3})--(\ref{PLIN3A}) we get
\begin{eqnarray}
\label{PLIN4}
&&r=16\sqrt{\frac{1-n}{1+n}}\left(\frac{1-n}{4-n}\right)(1-n_{S}).
\end{eqnarray}

Further, taking into account the observational constraint $r<0.032$ and restrictions Eqs. (\ref{GPL10})--(\ref{GPL11}) for the same observational values of spectral tilt $\tilde{n}_{S}=n_{S}=0.9663\pm 0.0041$. From Eq. (\ref{PLIN3}), we obtain the following constraints on the parametrization constant
\begin{eqnarray}
\label{PLIN5}
&&0.56<n<1,
\end{eqnarray}
corresponding to the verifiable inflationary models with potential Eq. (\ref{POTNEW3}).

Thus, from (\ref{WEND}) and (\ref{PLIN5}) we get the following estimation of the constant parameter $\lambda\simeq1/4$ in dependence (\ref{PLPM}).

Therefore, from (\ref{PLPM}) we obtain
\begin{equation}
\label{PLPML}
F(\phi(t))\simeq\left(4H(t)\right)^n.
\end{equation}
\color{black}

Also, we get the following possible values of the velocity of the scalar perturbations $0<c_{S}<0.53$ from expression (\ref{CS}). Thus, in the sub-luminal regime of the propagation of scalar perturbations,  chaotic inflation with quadratic potential corresponds to the observational constraints (\ref{PARCONSTRAINTNS})--(\ref{PARCONSTRAINTR}) and exit from inflation (\ref{INTCONDITION2}) as well.

Therefore, the super-luminal regime of propagation of scalar perturbations cannot be realized in this cosmological model.

\subsection{Correspondence between exact and approximate cosmological solutions}\label{SEC7.4}

The exact solutions of cosmological dynamic equations (\ref{PLPM})--(\ref{pole_3}) for Hubble parameter (\ref{INT6}) are
\begin{eqnarray}
\label{EXACT1}
&&\phi(t)=\frac{2}{2+n}\sqrt{\frac{2}{\mu}(1+n)}(\lambda-\mu t)^{1+\frac{n}{2}},\\
\label{EXACT2}
&&V(\phi)=V_{0}\phi^{\frac{2n}{2+n}}\left(\phi^{\frac{4}{2+n}}-B\right),\\
\label{EXACT2A}
&&F(\phi)=\frac{V_{0}}{3}\phi^{\frac{2n}{2+n}},
\end{eqnarray}
where
\begin{eqnarray}
\label{EXACT3}
&&V_{0}=\frac{3\mu(2+n)^{2}}{8\lambda^{n}(1+n)},\\
\label{EXACT4}
&&B=\frac{\mu}{3}(1+n)\left(\frac{8(1+n)}{\mu(2+n)^{2}}\right)^{\frac{2}{2+n}}.
\end{eqnarray}

As one can see, under condition $B\ll\phi^{\frac{4}{2+n}}$ exact form of potential (\ref{EXACT3}) corresponds to (\ref{POTNEW3B}) for the slow-roll approximation.

From (\ref{INT8}) we obtain following relation
\begin{eqnarray}
\label{EXACT5}
&&H=\left(\frac{\mu}{\epsilon}\right)^{1/2},
\end{eqnarray}
and at the crossing of the Hubble radius
\begin{eqnarray}
\label{EXACT5A}
&&H_{\ast}=\left(\frac{\mu}{\epsilon_{\ast}}\right)^{1/2}.
\end{eqnarray}

Based on (\ref{EXACT5}), we can rewrite the expression for the scalar field (\ref{EXACT1}) as follows
\begin{eqnarray}
\label{EXACT5B}
&&\phi(\epsilon)=\frac{2}{2+n}\sqrt{\frac{2}{\mu}(1+n)}\left(\frac{\mu}{\epsilon}\right)^{\frac{2+n}{4}}.
\end{eqnarray}

From (\ref{PLIN5}), (\ref{EXACT4}) and (\ref{EXACT5B}) we have
\begin{eqnarray}
\label{EXACT5C}
&&\frac{\phi^{\frac{4}{2+n}}}{B}=\frac{C}{\epsilon}\sim\epsilon^{-1},
\end{eqnarray}
where $1.5<C<1.9$.

Thus, condition $B\ll\phi^{\frac{4}{2+n}}$ is actual while slow-roll condition $\epsilon\ll1$ is satisfied.

Therefore, we can consider potential (\ref{EXACT2}) as
\begin{eqnarray}
\label{EXACT5D}
&&V(\phi)\sim\phi^{2},
\end{eqnarray}
during slow-roll regime.

At the end of inflation with $\epsilon=1$ the second term in potential (\ref{EXACT2}) becomes significant.

Also, from expression (\ref{PERTG3}) taking into account $\delta_{\ast}=0$, we obtain
\begin{eqnarray}
\label{EXACT6}
&&\epsilon_{\ast}=\frac{1-n_{S}}{4-n}.
\end{eqnarray}

From (\ref{PLIN5}) and (\ref{EXACT6}) for $n_{S}\simeq0.97$ we obtain the following values of the first slow-roll parameter at the crossing of the Hubble radius $0.0087<\epsilon_{\ast}<0.01$.
Therefore, at the crossing of the Hubble radius slow-roll conditions are satisfied.

From definitions of e-folds number implying $\dot{N}=H$ and the first slow-roll parameter (\ref{EMIN}) for Hubble parameter (\ref{INT6}) we get
\begin{eqnarray}
\label{EXACT7}
&&\frac{dN}{d\epsilon}=\frac{dN}{dt}\frac{dt}{d\epsilon}=\frac{H}{\dot{\epsilon}}=\frac{1}{2\epsilon^{2}}.
\end{eqnarray}

Thus, the e-folds number between the crossing of the Hubble radius and the end of inflation is
\begin{eqnarray}
\label{EXACT8}
\nonumber
&&\tilde{N}=\frac{1}{2}\int^{\epsilon_{e}=1}_{\epsilon_{\ast}}\frac{d\epsilon}{\epsilon^{2}}=
\frac{1}{2}\left(\frac{1}{\epsilon_{\ast}}-1\right)\\
&&\simeq \frac{1}{2\epsilon_{\ast}}=
\frac{4-n}{2(1-n_{S})}.
\end{eqnarray}

From (\ref{PLIN5}) and (\ref{EXACT8}) for $n_{S}\simeq0.97$ we obtain the following values of the e-folds number between the crossing of the Hubble radius and the end of inflation
\begin{eqnarray}
\label{EXACT9}
&&50<\tilde{N}<57.
\end{eqnarray}

Also, from (\ref{PERTG3})--(\ref{PERTG4}) we obtain the following expressions for the spectral index of the scalar perturbations and the tensor-to-scalar ratio
\begin{eqnarray}
\label{EXACT10}
&&1-n_{S}=\frac{4-n}{2\tilde{N}},\\
\label{EXACT11}
&&r=\frac{8(1-n)}{\tilde{N}}\sqrt{\frac{1-n}{1+n}},
\end{eqnarray}
which leads to the same dependence (\ref{PLIN4}).

From (\ref{PERTG2}) and (\ref{EXACT6}) we get
\begin{eqnarray}
\label{EXACT12}
&&{\mathcal P}_{S}=\frac{(2-n)(4-n)H^{2-n}_{\ast}}{8\pi^{2}(1-n^{2})\sqrt{\frac{1-n}{1+n}}\,(1-n_{S})}.
\end{eqnarray}

Thus, from (\ref{PLIN5}) and (\ref{EXACT12}) for ${\mathcal P}_{S}=2.1\times10^{-9}$ and $n_{S}\simeq0.97$ we obtain following scale of inflation
\begin{eqnarray}
\label{EXACT13}
&&0<H_{\ast}<3\times10^{-7},
\end{eqnarray}
in the chosen system of units.

Also, from (\ref{PLPML}) and (\ref{EXACT13}) we obtain the following constraint on the values of the coupling function at the crossing of the Hubble radius
\begin{eqnarray}
\label{EXACT13CF}
&&0<F_{\ast}<5\times10^{-4}.
\end{eqnarray}

From (\ref{PLIN5}), (\ref{EXACT5A}), (\ref{EXACT6}) and (\ref{EXACT13}) we obtain
\begin{eqnarray}
\label{EXACT14}
&&0<\mu<8\times10^{-16}.
\end{eqnarray}

From (\ref{WL4})--(\ref{PLPMCS}), (\ref{PLIN5}) and (\ref{EXACT3})--(\ref{EXACT4}) we get
\begin{eqnarray}
\label{EXACT15V}
&&V_{0}<5.4\times 10^{-15},\\
\label{EXACT16B}
&&B<4.2\times 10^{-4}.
\end{eqnarray}

Also, from (\ref{POT}), (\ref{EXACT13CF}) and (\ref{EXACT14}) the value of the potential of the scalar field  at the crossing of the Hubble radius is
\begin{eqnarray}
\label{POTE}
&&V_{\ast}=F_{\ast}\left(3H^{2}_{\ast}-\mu(1+n)\right)<1.3\times10^{-16}.
\end{eqnarray}

Kinetic energy of the scalar field (\ref{FIELD}) at the crossing of the Hubble radius is
\begin{eqnarray}
\label{KINX}
&&X_{\ast}=F_{\ast}\mu(1+n)<8\times10^{-19}.
\end{eqnarray}

Thus, the slow-roll condition $V_{\ast}\gg X_{\ast}$ is satisfied in this inflationary model.

In natural system of units~\cite{Baumann:2014nda} potential (\ref{POTE}) can be estimated as follows
\begin{eqnarray}
\label{POTE2}
&&V_{\ast}<1.36\times10^{-16}\times M^{4}_{P}=4.3\times10^{57}~(GeV)^{4},
\end{eqnarray}
where reduced Planck mass is $M_{P}=\left(8\pi G\right)^{-1/2}=2.4\times10^{18}$ GeV.

Corresponding energy scale of inflation is
\begin{eqnarray}
\label{POTE3}
&&V^{1/4}_{\ast}<2.6\times10^{14}~GeV<\Lambda_{GUT}<M_{P},
\end{eqnarray}
where $\Lambda_{GUT}\sim 10^{16}$ GeV is GUT scale which is considered as the upper bound on the energy scale of inflationary models~\cite{Baumann:2014nda}.
\color{black}

Also, from (\ref{PLIN5}), (\ref{EXACT5B}), (\ref{EXACT6}) and (\ref{EXACT14}) we have following estimates for the value of a scalar field
\begin{eqnarray}
\label{EXACT16}
&&0<\phi_{\ast}<0.22,\\
\label{EXACT17}
&&0<\phi_{e}<0.01.
\end{eqnarray}

Now,  we will check to what extent this model satisfies Swampland criteria~\cite{Odintsov:2023weg,Das:2018hqy,Oikonomou:2023bmn}
\begin{eqnarray}
\label{EXACT18}
&&\left|\Delta\phi\right|=\left|\phi_{e}-\phi_{\ast}\right|\leq{\mathcal O}\left(1\right),\\
\label{EXACT19}
&&\left|\frac{V_{,\phi}}{V}\right|_{\phi=\phi_{\ast}}\geq {\mathcal O}\left(1\right).
\end{eqnarray}

Satisfaction of the Swampland criteria means that the cosmological model can be regarded as a low-energy effective one in quantum gravity~\cite{Odintsov:2023weg,Das:2018hqy,Oikonomou:2023bmn}.

On the basis of expressions (\ref{EXACT5D}) and (\ref{EXACT16})--(\ref{EXACT17}) we obtain
\begin{eqnarray}
\label{EXACT20}
&&\left|\Delta\phi\right|<0.21<{\mathcal O}\left(1\right),\\
\label{EXACT21}
&&\left|\frac{V_{,\phi}}{V}\right|_{\phi=\phi_{\ast}}=\frac{1}{\phi_{\ast}}\simeq4.5>{\mathcal O}\left(1\right).
\end{eqnarray}

Thus, the proposed model of cosmological inflation based on the scalar-torsion gravity with potential (\ref{EXACT2}) and coupling function (\ref{EXACT2A}) satisfies the Swampland criteria, observational constraints on the parameters of cosmological perturbations and the condition of the graceful exit from inflation.

\section{Conclusion}\label{SEC8}
We testify the influence of non-minimal coupling between scalar field and torsion on the cosmological parameters based on power-law relationship between coupling function and Hubble parameter $F=(H/\lambda)^{n}$. The novelty of the proposed approach is the possibility of a model-independent assessment of non-minimal coupling effect on cosmological parameters.
The non-minimal coupling influence was parameterized by one constant parameter $n$. Also, we found restrictions on the value of parametrization constant $-1<n<1$. The influence of non-minimal coupling on cosmological dynamics, scalar field potential, and cosmological perturbations parameters was assessed by this approach.

We have shown that inflationary models implying a linear relationship between the tensor-scalar ratio and spectral index of scalar perturbations can be verified by observational constraints on the values of cosmological perturbation parameters by influence of the non-minimal coupling between scalar field and torsion without the need to take into account the Galilean field self-interaction. These influence leads to a sub-luminal regime of the propagation of scalar perturbations and a decrease in the value of the tensor-scalar ratio for inflationary models under consideration.
A non-trivial property of the power-law relationship $F=(H/\lambda)^{n}$ is the constant velocity of propagation of scalar perturbations in inflationary models under consideration. Also, we note that expressions (\ref{PLPM}) and (\ref{HREL}) define the relations of the constant parameters $n$ and $\lambda$ with $c_{S}$.
Thus, in this case, the parametrization of the influence of non-minimal coupling between the scalar field and torsion can be performed only based on the propagation velocity of scalar perturbations $c_{S}=constant$ following expressions (\ref{WL4})--(\ref{PLPMCS}).

Further development of the proposed approach is associated with its generalization to the case of the violation of relations $H=h^{1/(1+n)}$ and  $\delta/\epsilon=const$.
This generalization leads to a wider class of cosmological models beyond the considered type of inflationary model.

Finally, we note that taking into account the Galilean field self-interaction can also significantly increase the number of the possible verifiable inflationary models.

\section*{Acknowledgements} BM acknowledges Bauman Moscow State Technical University(BMSTU), Moscow for providing the Visiting Professor position during which part of this work has been accomplished. LKD acknowledges the financial support provided by University Grants Commission (UGC) through Senior Research Fellowship UGC Ref. No.: 191620180688 to carry out the research work.

\section*{References}
\bibliographystyle{utphys}
\bibliography{references}

\end{document}